\documentclass[sigconf]{acmart}
\AtBeginDocument{%
  \providecommand\BibTeX{{%
    \normalfont B\kern-0.5em{\scshape i\kern-0.25em b}\kern-0.8em\TeX}}}

\setcopyright{acmcopyright}
\copyrightyear{2021}
\acmYear{2021}
\acmDOI{10.****/*****.*******}
\acmConference[St. Louis '21]{St. Louis '21: The International Conference for High Performance Computing, Networking, Storage, and Analysis}{November 14--19, 2021}{St. Louis, MO}
\acmBooktitle{St. Louis '21: The International Conference for High Performance Computing, Networking, Storage, and Analysis,
  November 14--19, 2021, St. Louis, MO}
\acmPrice{15.00}
\acmISBN{978-1-4503-XXXX-X/18/06}

\begin{document}

\title{Datastore Design for Analysis of Police Broadcast Audio at Scale}

\author{Ayah Ahmad}
\orcid{0000-0002-4481-866X}
\affiliation{%
  \institution{University of California, Berkeley}
  \textit{Department of Electrical Engineering \& Computer Sciences}
  \\
  \city{Berkeley}
  \state{California}
  \country{USA}}
\email{ayahahmad@berkeley.edu}

\author{Christopher Graziul (advisor)}
\affiliation{%
  \institution{University of Chicago}
  \textit{Department of Comparative Human Development}
  \\
  \city{Chicago}
  \state{Illinois}
  \country{USA}}
\email{graziul@uchicago.edu}

\author{Margaret Beale Spencer (advisor)}
\affiliation{%
  \institution{University of Chicago}
  \textit{Department of Comparative Human Development}
  \\
  \city{Chicago}
  \state{Illinois}
  \country{USA}
}
\email{mbspencer@uchicago.edu}

\renewcommand{\shortauthors}{Ahmad, et al.}

\begin{abstract}
With policing coming under greater scrutiny in recent years, researchers have begun to more thoroughly study the effects of contact between police and minority communities. Despite data archives of hundreds of thousands of recorded Broadcast Police Communications (BPC) being openly available to the public, a closer look at a large-scale analysis of the language of policing has remained largely unexplored. While this research is critical in understanding a "pre-reflective" notion of policing, the large quantity of data presents numerous challenges in its organization and analysis.

In this paper, we describe preliminary work towards enabling Speech Emotion Recognition (SER) in an analysis of the Chicago Police Department's (CPD) BPC by demonstrating the pipelined creation of a datastore to enable a multimodal analysis of composed raw audio files.

\end{abstract}




\maketitle

\section{Introduction}
In this section, we discuss the data that we operate on, relevant literature that influenced datastore design choices, and the framework that forms the foundation for this project.

\subsection{Data}
The data we operated on consisted of a public archive of 160,000 30-minute audio files of the CPD's BPC. Each audio file is 30 minutes long and approximately 3.5 MB, totaling approximately 4.8 million minutes and 560 GB for the raw archive. Associated with each audio file was metadata extracted from the name of the file, including dispatch zone and date, and metadata extracted using Voice Activity Detection (VAD), including timestamps of non-silent slices of audio.
\subsection{Literature Review}
To survey the research in Speech Emotion Recognition, we relied heavily on \cite{Akcay20} to contextualize and summarize pertinent SER models. In this search, we sought out prevalent research that focused on, or was bound by, the following constraints: 
\begin{enumerate}
    \item Examined \textbf{elicited emotions} during improvised conversations, as opposed to acted emotions
    \item Was \textbf{robust to noise}, particularly at a level close to human speech
    \item Used a \textbf{3-dimensional model of emotion}, instead of a categorical model
\end{enumerate}
These constraints fundamentally allowed us to look closer at the research more applicable to the data we were analyzing. While no model that we examined fulfilled all three requirements, \cite{8421023} was perhaps the closest. There, they created a 3-D Convolutional Recurrent Neural Network (CRNN) for SER. To avoid adding potential bias, we decided to extract audio features and use those as inputs, instead of using human-labeled data.
\subsection{PVEST Framework}
The Phenomenological Variant of Ecological Systems Theory \cite{Spencer07} serves as the theoretical framework supporting this project. In this context, the framework seeks to identify adaptive and maladaptive coping mechanisms of police officers' stress responses. Thus, we seek to understand how normal communication can contribute to increasing or decreasing the likelihood of an adverse encounter between police and the general public—and more specifically, Law Enforcement Officers (LEOs) and Male Minority Youth (MMY).

\section{Challenges}
This section will discuss challenges associated with the creation of a database, due to the scale, temporality, and silence of the data.
\subsection{Scale}
In expanding the audio into discrete samples, we ended up with approximately 40 million data points. From there, we extracted 26 temporal features, at approximately 183,000 samples per feature—based on the Geneva Minimalistic Acoustic Parameter Set (GeMAPS) \cite{60dffd1c51834c7c996272c419934d7c}—using openSMILE \cite{10.1145/1873951.1874246}, resulting in approximately 690,000 data points per file. Using Praat-Parselmouth \cite{JADOUL20181, boersma2021a} to extract intensity, harmonicity, and pitch for each file resulted in approximately 230,000 data points per feature, per file. Scaling upwards, to include all 160,000 audio files results in approximately 7.2  trillion data points for the raw audio, GeMAPS, and Parselmouth files. 

\subsection{Temporality}
When extracting different features from individual files, the default sample rate varies from one program to another. Thus, for each audio file, we have both raw audio data for each 22kHz sampling period and features extracted at differing periods of time. 

\subsection{Silence}
Since some files contain silent slices, clustering on data that contains silence could lead to an inherently binary model, explained by one dimension—silence or sound.

\section{Database Design and Implementation}
In searching for a database management system (DBMS) that was scalable, and ACID-compliant, extensible, with high levels of concurrency, we decided to use PostgreSQL. 

Operating under a 1TB constraint meant that we could not store our raw and extracted data directly in the database because this composition of features exceeded 30 TB. This was in addition to the constraint set by the misalignment in temporality. Thus, we determined to design a datastore, such that raw and extracted features could simultaneously be accessed and preprocessed as inputs to a CRNN. Each file is stored in a specified location, with the locations used instead of the files in the database. Thus, for any feature stored as a column in the database, a script would extract the file locations, and feed them into a clustering algorithm that would parse the given file and cluster the data accordingly. Similarly, for the SER model, a script would perform the same parsing of the files and use the parsed data as inputs to the model.

\section{Conclusion}
In this project, we created a framework that enabled easy interoperability with statistical methods for an unbiased large-scale analysis of police broadcast audio for SER, allowing us to do large-scale pre-processing, clustering and PCA on the dataset.

\begin{acks}
Research reported in this publication was supported by the National Institute On Minority Health And Health Disparities of the National Institutes of Health under Award Number R01MD015064. 
\end{acks}

\bibliographystyle{ACM-Reference-Format}
\bibliography{sample-base}


\end{document}